\documentclass{epl}

\title{Stochastic dynamics of adhesion catch-slip bond cluster}
\author{F. Liu\inst{1} \and Z.-C. Ou-Yang\inst{1,2}}

\institute{
  \inst{1} Center for Advanced Study, Tsinghua University, Beijing, China\\
  \inst{2} Institute of Theoretical Physics,
The Chinese Academy of Sciences, P.O.Box 2735 Beijing 100080,
China }

\begin{document}

\maketitle

\begin{abstract}
We present a general stochastic dynamic model of a cluster of
biological complexes with fluctuating dissociation and association
rates. The master equation has analytical solutions in two
limiting cases: the vanishing force with reflecting boundary
condition and slow reaction limit. In particular, the latter can
describe the dynamics of a catch-slip bond cluster under a
constant force. We find that if the rebinding rate vanishes, the
size of the cluster matters little to the cluster lifetime within
the whole catch force regime. The large cluster rapidly decays and
single catch bond governs the final dynamic behaviors of the
cluster. The catch behavior would be further amplified in the
presence of large nonvanishing rebinding rates.
\end{abstract}

\section{Introduction}

Recent experiments found that at single molecule
level~\cite{Marshall,Evans04,Sarangapani}, the dissociation rates
of some biological adhesion complex, including the bonds between
L- and P-selectin and P-selectin glycoprotein ligand 1 (PSGL-1)
firstly decrease with initial application of small force, and then
increase when the force increases beyond some threshold. This
counterintuitive behavior has been termed catch-slip bond
transitions. Because these bonds are primarily responsible for the
tethering and rolling of leukocytes on inflamed endothelium under
shear stress~\cite{McEver,Konstantopoulos98}, this discovery may
provide a direct experimental evidence at the single-molecule
level to account for the shear threshold
effect~\cite{Finger,Lawrence}, in which the number of rolling
leukocytes first increases and then decreases while monotonically
increasing shear stress. Several kinetic
models~\cite{Evans04,Barsegov,Pereverzev} have been developed to
explain the catch-slip bond behavior at the single molecule level.
We~\cite{liufPRE,liufSUB1,liufSUB2} recently suggested alternative
mechanism. We proposed that the forced dissociations of
PSGL-1$-$L- and $-$P might a typical rate process with dynamic
disorder~\cite{Zwanzig}. The catch-slip bond transitions can be
induced by a particular energy barrier shape with respect to the
inner conformational coordinate~\cite{liufSUB1}. Although applied
force projection on the dissociation reaction only accelerates the
dissociations, the component of the force along the conformational
coordinate could stabilize the molecular complex by dragging the
molecule to higher energy barriers; if effect of the latter is
greater than that of the former, the bond presents catch behavior,
otherwise a slip bond is observed

But before we apply the concept of the catch-slip bond transitions
to the real problems, we must know that, in physiological setting
adhesion complexes usually exist in a cluster and cooperatively
realize their functions with clusters~\cite{Bell}. Two new
features will emerge in a cluster of bonds. Firstly the effect of
bond association has to be considered. For single bond case, in
general this effect could be neglected due to elastic recoil of
the force transducer after complex rupture~\cite{Evans01}. In
contrast, the dissociated complex can be formed again when the
other bonds are still closed. On the other hand, because the
closed complexes share the same external force, any dissociation
or association of one bond would alter the partition of the force
on the other closed bonds. It means that we must consider the
cooperativity between bonds in clusters. There are considerable
theoretical works contributed to the behaviors of a cluster of
slip
bonds~\cite{Bell,Dembo,Seifert00,Seifert02,Lipowsky,Tees,Cozens,
Erdmann04PRL,Erdmann04EL,Erdmann04JCP}. We naturally ask what
difference will be observed between a cluster of catch-slip bonds
and slip bonds. We try to give an answer in the present work.
Considering that the dissociation of a catch-slip bond is a rate
process with dynamic disorder, we begin with a general stochastic
dynamical model of a cluster of molecular complex with fluctuating
dissociation and association rates.

\section{Model} Following the previous theoretical work~\cite{liufSUB1},
we assume that the forced dissociation rate of single bond has the
Bell expression $k^{+}_f(x)=k^{+}(x)\exp(\beta f\xi^\ddag)$, where
$x$ is the complex conformational coordinate~\cite{Agmon},
$\beta=k_{\rm B}T$, $k_{\rm B}$ is the Boltzmann's constant, $T$
is absolute temperature, $\xi^\ddag$ is a projection of the
distance from the bound state to the energy barrier onto the
external force $f$, $k^{+}(x)$ is the intrinsic dissociation rate
and has Arrhenius form, i.e., $k^{+}(x)\propto\exp\left[\beta
\Delta G^\ddag(x) \right]$. The rebinding rate $k^{-}(x)$ is
independent of force~\cite{Bell,Seifert00,Erdmann04JCP}. Given a
cluster of $N$ parallel catch-slip bonds, if the closed bonds
sharing the constant force equally, the probability that $i$ bonds
are closed and has particular conformational coordinate $x_i$ at
time $t$ then satisfies the following one-step master equation
\begin{eqnarray}
\label{masterequation}
\frac{\partial}{\partial t}P_0(x^0_1\cdots
x^0_N,t)&=&\left[\sum_i{\cal L}^0(x_i)\right]P_0(x^0_1\cdots
x^0_N,t)+\sum_i k^{+}_f(x_i)P_1(x^0_1\cdots x^1_i\cdots
x^0_N,t),\nonumber\\
\frac{\partial}{\partial t}P_1(x^0_1\cdots x^1_i\cdots x^0_N,t)&=&
\left[{\cal L}^1_f(x_i)+\sum_{j\neq i}{\cal L}^0(x_j)-k^+_f(x_i)-
\sum_{j\neq i}
k^-(x_j)\right]P_1(x^0_1\cdots x^1_i\cdots x^0_N,t)\nonumber\\
&{}&+\sum_{j\neq i} k^+_{f/2}(x_j)P_2(x^0_1\cdots x^1_j\cdots
x^1_i\cdots x^0_N,t),\nonumber\\
\frac{\partial}{\partial t}P_m(x^{\sigma_1}_1\cdots
x^{\sigma_N}_N,t)&= &\left[\sum_{b_i}{\cal
L}^1_{f/m}(x_{b_i})+\sum_{u_i}{\cal L}^0(x_{u_j})-\sum_{b_i}
k^+_{f/m}(x_{b_i}) \right. \\
&&\left.-\sum_{u_i}
k^-(x_{u_i})\right]P_m(x^{\sigma_1}_1\cdots x^{\sigma_N}_N,t)\nonumber\\
&&+\sum_{u_i} k^+_{f/m+1}(x_{u_i})P_{m+1}(x^{\sigma_1}_1\cdots
x^1_{u_i}\cdots x^{\sigma_N}_N,t)\nonumber\\
&&+\sum_{b_i}  k^-(x_{b_i})P_{m-1}(x^{\sigma_1}_1\cdots
x^0_{b_i}\cdots x^{\sigma_N}_N,t),\nonumber\\
\frac{\partial}{\partial t}P_N(x^1_1\cdots
x^1_N,t)&=&\left[\sum_i{\cal L}^1_{f/N}(x_i)-\sum_i
k^+_{f/N}(x_i)\right]P_N(x^1_1\cdots x^1_N,t)\nonumber\\
&&+\sum_i k^{-}(x_i)P_{N-1}(x^1_1\cdots x^0_i\cdots
x^1_N,t)\nonumber
\end{eqnarray}
where $u_i$ and $b_i$ respectively denote the ruptured and closed
bond number, the superscript $\sigma_i=1$, $0$ respectively
represents $i$-bond ruptured or not, and $\sum_i^{N}\sigma_i=m$,
$2\le m\le N-1$; the Fokker-Planck operators are defined by
\begin{eqnarray}
{\cal L}^1_{f}(x)=D\frac{\partial}{\partial x}e^{-\beta
V_{f}(x)}\frac{\partial}{\partial x}e^{\beta V_{f}(x)}, \textrm{
and } {\cal L}^0(x)=D\frac{\partial}{\partial x}e^{-\beta
U_0(x)}\frac{\partial}{\partial x}e^{\beta U_0(x)},
\end{eqnarray}
where $D$ is the diffusion coefficient, and the potential
$V_{f}(x)=U_1(x)-fx$~\cite{liufSUB1}. Because rebinding of the
completely dissociated state is usually prevented by elastic
recoil of the transducer, we write the first two equations (m=0,
1) independently. According to previous theoretical
work~\cite{Erdmann04JCP}, they were termed as the absorbing
boundary conditions of the master equation at $m=0$ and 1. If at
initial time, all bonds are closed and in thermal equilibrium,
i.e.,
\begin{eqnarray}
P_N(x^1_1,\cdots x^1_N,0)=\prod_{i=1}^N P^1_{\rm eq}(x_i),
\end{eqnarray} and others vanish. where the distribution
$P^1_{\rm eq}(x_i)\propto \exp\left[-\beta U_1(x)\right]$, we'd
like to calculate the cluster lifetime, $T_f=\int_0^\infty t
D_f(t)dt $. Here
\begin{eqnarray}
D_f(t)=\int dP_0(x^0_1\cdots x^0_N,t)/dt \prod_{i=1}^Ndx_i
\end{eqnarray}
is an extended cluster dissociation rate with
dynamic disorder. We see this master equation are far complicated
than those in the absence of dynamic disorder. It is impossible to
analytically solve eq.~(\ref{masterequation}) for general rate
functions and potentials with respect to the conformational
coordinate. Numerical approach such as Monte Carlo
simulations~\cite{Erdmann04JCP} is also not simple to be applied
in the current case. In the following sections, we consider only
two limiting cases, the vanishing force with reflecting boundary
conditions and slow reaction limits, in which the master equation
has analytical solutions.
 \section{Vanishing force with reflecting boundary
conditions} In this case, the reflecting boundary
condition~\cite{Erdmann04JCP} means that the first two equations
in eq.~(\ref{masterequation}) are altered into,
\begin{eqnarray}
\label{stochasticdynamicsvanishingforce} \frac{\partial}{\partial
t}P_0(x^0_1\cdots x^0_N,t)&=&\left[\sum_i{\cal L}^0(x_i)-\sum_i
k^-(x_i)\right]P_0(x^0_1\cdots x^0_N,t)\nonumber\\
&& +\sum_i
k^{+}(x_i)P_1(x^0_1\cdots x^1_i\cdots x^0_N,t),\nonumber \\
\frac{\partial}{\partial t}P_1(x^0_1\cdots x^1_i\cdots x^0_N,t)&=&
\left[{\cal L}^1(x_i)+\sum_{j\neq i}{\cal
L}^0(x_j)-k^+(x_i)\right.\nonumber\\
&&\left.-\sum_{j\neq i}
k^-(x_j)\right]P_1(x^0_1\cdots x^1_i\cdots x^0_N,t)\nonumber\\
&+&\sum_{j\neq i} k^+(x_j)P_2(x^0_1\cdots x^1_j\cdots x^1_i\cdots
x^0_N,t)\nonumber\\
&&+k^-(x_i)P_0(x^0_1\cdots x^0_i\cdots x^0_N,t).\nonumber
\end{eqnarray}
The other terms ($m\ge 2$)are the same with those in
eq.~(\ref{masterequation}) except the force therein vanishing. It
is easy to prove that $P_m(x^{\sigma_1}_1\cdots
x^{\sigma_N}_N,t)=\prod^N_1P^{\sigma_i}(x_i,t)$ is the solutio in
this limit, where $P^{\sigma}(x,t)$, $\sigma=0,1$ are solutions of
the dissociation-association diffusion equations of single bond in
the absence of force given by
\begin{eqnarray}
\label{reversereactiondiffusionequation} \frac{\partial
P^1(x,t)}{\partial
t}&=&{\cal L}^1(x)P^1(x,t)- k^+(x)P^1(x,t)+k^-(x) P^0(x,t),\nonumber\\
\frac{\partial P^0(x,t)}{\partial t}&=&{\cal L}^0(x)P^0(x,t)
+k^+(x)P^1(x,t)-k^-(x)P^0(x,t).
\end{eqnarray}
Here the initial conditions are $P^1(x,0)\propto \exp(-\beta
U_1(x))$ and $P^0(x,0)=0$. Therefore, any quantity of interest can
now be calculated. For instance, the probability $Q_m(t)$ that
$m$-bond are closed at time $t$ is
\begin{eqnarray}
\label{probabilitymbondsatt} Q_m(t)&=&\sum_{
\sigma_1\cdots\sigma_N}\int P_m(x^{\sigma_1}_1\cdots
x^{\sigma_N}_N,t)
\prod^N_1 dx_i\nonumber\\
&=& C^m_N\left(\int P^1(x,t)dx\right)^m\left(\int
P^0(x,t)dx\right)^{N-m}.
\end{eqnarray}
Although the master equation with reflecting boundary conditions
has been formally solved, it is still a great challenge to solve
eq.~(\ref{reversereactiondiffusionequation}) analytically. Various
approximation approaches have been developed in the past two
decades~\cite{Portman}. We are not ready to further discuss it.
Instead, we study the other limit, in which is relevant to the
dynamics of a catch-slip bond cluster.
\section{Slow reaction limit}
In this limit the processes of the rupture and rebinding of any
single bonds are very slow compared to the conformational
diffusions. The thermal equilibrium distribution of the
conformation for each bond hence is always maintained during the
courses of ``reactions". We assume
\begin{eqnarray}
P_m(x^{\sigma_1}_1\cdots
x^{\sigma_N}_N,t)=q_m(\sigma_1\cdots\sigma_N,t)\prod_{b_i}
P^{1}_{\rm eq}\left(x_{b_i},f/m\right)\prod_{u_i} P^{0}_{\rm
eq}(x_{u_i}),
\end{eqnarray}
where $P^{\sigma}_{\rm eq}$, $\sigma=0,1$ are the thermal
equilibrium distributions and satisfy
\begin{eqnarray}
{\cal L}^1_{f/m}(x_{b_i})P^1_{\rm eq}\left(x_{b_i},f/m\right)=0,
\textrm{ and } {\cal L}^0(x_{u_i})P^0_{\rm eq}(x_{u_i})=0,
\end{eqnarray}
respectively. Substituting in eq.~(\ref{masterequation}) and
integrating all conformational freedoms lead to a master equation
about $q_m(\sigma_1\cdots\sigma_N,t)$. According to the definition
of $Q_m(t)$ (eq.~(\ref{probabilitymbondsatt})), we can further
simplify the new master equation as follows,
\begin{eqnarray}
\label{masterequationrapiddiffusion}
\frac{d}{dt}Q_0(t)&=&K^+_1(f)Q_1(t)\nonumber\\
\frac{d}{dt}Q_1(t)&=&-\left[K^+_1(f)+K^-_1\right]Q_1(t)+K^+_2Q_2(t)\nonumber\\
\frac{d}{dt}Q_m(t)&=&-\left[K^+_m(f)+K^-_m\right]Q_m(t)-K^+_{m+1}(f)Q_{m+1}(t)+
K^-_{m-1}Q_{m-1}(t)\\
\frac{d}{dt}Q_N(t)&=&-K^+_N(f)Q_N(t)+K^-_{N-1}Q_{N-1}(t)\nonumber
\end{eqnarray}
where
\begin{eqnarray}
&&K^+_m(f)=me^{\beta f\xi^\ddag/m}\int k^+(x)P^{1}_{\rm
eq}(x,f/m)dx, \nonumber\\
&&K^-_m=(N-m)\int k^-(x)P^0_{\rm eq}(x)dx=(N-m)\langle k^-\rangle.
\end{eqnarray}
We see eq.~(\ref{masterequationrapiddiffusion}) is formally the
same with the one-step master equation that describes the force
ruptured dynamics of a cluster of slip bond~\cite{Erdmann04JCP}.
Many analytical results that are independent of concrete
expression of the rates hence could be applied here. On the other
hand, because the new defined rate constant $K_m^+(f)$ is a
dynamical average of the Bell expression on the distribution
$P_{\rm eq}(x,f/m)$, its force dependence could be very different
from that of conventional slip bond. Particularly, if the
potential $U_1(x)$ is harmonic with spring constant $\kappa$, and
barrier $\Delta G^\ddag(x)$ is a piecewise function with two
segments, $K_m^+(f)$ is approximated to be
\begin{displaymath}
 K_m^+(f) \approx \left\{ \begin{array}{ll}
  mk_c \exp(-\beta d^\ddag_c f/m) & f<f_t \\
  mk_s\exp(+\beta d^\ddag_s f/m) & f>f_t, \\
\end{array}  \right.
\end{displaymath}
where $f_t$ is the bond transition force, $k_i$, $i=d,s$ are some
constants, and the ``effective" distance $d^\ddag_i>0$, are the
simple combinations of $\kappa$, the slopes of segments, and
$\xi^\ddag$~\cite{liufSUB1}. We see if force is smaller than
$f_t$, the rupture rate $K_m^+(f)$ decreases with increasing force
(a catch bond cluster); otherwise it is a slip bond cluster. For
the latter, eq.~(\ref{masterequationrapiddiffusion}) is completely
the same with previous models~\cite{Erdmann04JCP,Erdmann04PRL}.
Hence we only study the dynamics of the catch bond cluster. We
respectively discuss two cases according to zero and nonzero
rebinding rates.
\subsection{Rebinding rates $k^{-}(x)=0$} In this case, the
probability $Q_m(t)$ of m-bond closed at time $t$ has an exact
solution given by~\cite{Erdmann04PRL}
\begin{eqnarray}
Q_m(t)=\left(\prod^N_{n=m+1} K^+_{n}(f)
\right)\sum^N_{n=m}\left[e^{-tK^+_{n}(f)}\prod^N_{p=m\atop p\neq
n}\left(K^+_{p}(f)-K^+_{n}(f)\right)^{-1}\right]
\end{eqnarray}
Hence, any quantity of interest can be calculated from $Q_m(t)$.
For instance, the cluster lifetime $T_f$ is
\begin{eqnarray}
\label{averagelifetime} T_f=\sum\limits_{m=1}^N {K^+_m(f)}^{-1}.
\end{eqnarray}
For large $N$ and small force $f<f_c=k_{\rm B}T/d_c$, the above
equation can be approximated to be~\cite{Goldstein}
\begin{eqnarray}
T_f\approx H_N\approx \ln N+1/2N+0.577,
\end{eqnarray}
where $H_N=\sum_i^N i^{-1}$ is the harmonic number. It is same
with the lifetime of slip bond cluster~\cite{Erdmann04JCP}. It is
expected because at small forces the difference between the catch
and slip bonds are negligible. If $f>f_c$, the force in the
exponentials of $T_f$ cannot be neglected again. In contrast, for
the larger force the dominating term in
eq.~(\ref{averagelifetime}) is the one for the rupture of only one
bond, i.e.,
\begin{eqnarray}
T_f\sim {k_c}^{-1}\exp(f/f_c).
\end{eqnarray}
This result is further demonstrated by the time decay function of
the mean number of the closed bonds on large time scales:
\begin{eqnarray}
\langle m\rangle\sim\exp\left[{-t k_c\exp(-f/f_c)}\right ].
\end{eqnarray}
Compared to the behaviors of a slip bond cluster with large size,
three difference are observed in a large size of catch bond
cluster if the rebinding rate vanishes. First, the size of the
catch bond cluster matters little for the cluster lifetime within
the whole catch force regime, whereas it only holds for smaller
forces in the slip bond cluster. Then, for larger force the
dynamics of the catch cluster is took over by single catch bond.
In contrast, the largest clusters dominate the decay of the slip
bond cluster~\cite{Erdmann04JCP}. Finally, as a characteristic of
a catch bond, the increasing force only exponentially elongates
the lifetime of the catch bond cluster.

\subsection{Nonzero rebinding rates}
Although $Q_m(t)$ in Eq.~(\ref{masterequationrapiddiffusion}) in
principle could be calculated by eigenvalue technique in this
case, except the simplest case $N=2$, it is not more efficient
than the exact stochastic simulations~\cite{Erdmann04JCP}.
Fortunately, it has been proved~\cite{Erdmann04JCP,Kampen} that
the cluster lifetime of great interest has still an exact form
\begin{eqnarray}
\label{averagelifetimerebinding} T_f=\sum\limits_{m=1}^N
{K^+_m(f)}^{-1} + \sum\limits_{m = 1}^{N-1} \sum\limits_{n=m +
1}^N \left(\prod\limits_{p=n-m}^{n-1} K_p^- \left/
\prod\limits_{p=n-m}^n K_p^+(f) \right.\right).
\end{eqnarray}
We are not interested in the general dependence of
eq.~(\ref{averagelifetimerebinding}) on force and $N$. Instead we
consider a specific case where the ratio $\gamma=\langle
k^-\rangle/k_c$ is larger than 1. Thus for large $N$ and force,
eq.~(\ref{averagelifetimerebinding}) is
\begin{eqnarray}
T_f  \sim {k_c}^{-1}e^{f/f_c}\left[ 1 + \frac{1}{N}\gamma^{N -1}
e^{f(H_N-1)/f_c}\right].
\end{eqnarray}
In contrast to the case without rebinding, the size of the catch
cluster will change the cluster lifetime dramatically as
$\gamma>1$; the presence of the rebinding rate further amplifies
the lifetime by $\gamma^{N-1}$-fold. Therefore, the effect of the
rebinding are the same for the slip and catch bond cluster.

\section{Discussion and conclusions}
In this work, we present a general stochastic dynamic model of a
cluster of biological complexes with dynamic disorder. We study
two simplest limiting cases: vanishing force with reflecting
boundary conditions and the slow reaction limits. Under the former
limit, because the dissociation and association of each bonds are
independent, the solution to the model is a simple linear
combination of the dynamics of single bond. Under the slow
reaction limit, we demonstrate that the model can be reduced to
the conventional one-step master equation except that the rates
are replaced by dynamic averages. Particularly, given a specific
energy barrier with respect to molecular conformational
coordinate, this master equation can describe the dynamics of a
cluster of catch-slip bonds. We found that many conclusions
obtained for the slip bond cluster could be applied to the catch
bond cluster with minor modification. We give simple relations
between the cluster lifetime and the cluster size N, force $f$ and
the average rebinding rate. These conclusion would be useful in
future dynamic studies of the catch-slip bond clusters.

\acknowledgments One of the authors (F.L.) would like to thank
Prof. Seifert, U. for providing his interesting works to us. F.L.
was supported by the National Natural Science Foundation of China
(NSFC).

\end{document}